\documentclass[useAMS,usenatbib]{mn2e}
\usepackage{epsfig}
\usepackage{graphicx}
\usepackage{hyperref}
\begin{document}

\title[On the Possibility of the Detection of Extinct Radio Pulsars]
{On the Possibility of the Detection of Extinct Radio Pulsars}

\author[V.S. Beskin, S.A. Eliseeva]{V.S. Beskin$^1$ and S.A.Eliseeva$^2$
\thanks{E-mail: sa\_eliseeva@comcast.net; beskin@lpi.ru}\\
\thanks{Astronomy Letters, 2003, {\bf 29}, (1), 20-25, Translated from Russian by V. Astakhov. Corrected version.}\\
$^1$ P.N.~Lebedev Physical Institute, Leninsky prosp. 53, Moscow, 119991, Russia
\\
$^2$ Moscow Institute of Physics and Technology, Dolgoprudny, Moscow region, 141700, Russia}

\pagerange{\pageref{firstpage}--\pageref{lastpage}} \pubyear{2002}

\maketitle

\label{firstpage}

\begin{abstract}
We explore the possibilities for detecting pulsars that have ceased to radiate in the radio band.
We consider two models: the model with hindered particle escape from the pulsar surface [first suggested
by \citealt{RS1975}] and the model with free particle escape (\citealt{Arons1981}; \citealt{Mestel1999}).
In the model with hindered particle escape, the number of particles that leave the pulsar magnetosphere
is small and their radiation cannot be detected with currently available instruments. At the same time,
for the free particle escape model, both the number of particles and the radiation intensity are high enough for such pulsars to be detectable with the presently available receivers such as GLAST and AGILE spacecrafts. It is also possible that extinct radio pulsars can be among the unidentified EGRET sources.
\end{abstract}

\section{Introduction}
The number of radio pulsars in the modern catalogues already exceeds 1500 \citep{atnf} and it is still increasing. Most of the observed pulsars belong to young neutron stars with the characteristic ages of $\tau_D \sim 10$ Myr. Since the lifetimes of radio pulsars are much shorter than the age of the Universe, up to $10^9$ older neutron stars that have already ceased to radiate in radio band can exist in our Galaxy \citep{Lipunov1987}. The possibility of their detection is usually associated with the thermal radiation due either to internal stores of neutron-star thermal energy \citep{Yakovlev1999} or to accretion from the interstellar gas \citep{Colpi1998}. Either way the electrodynamic processes in the neutron-star magnetosphere are assumed to cease to play a crucial role. However, the cessation of secondary-plasma production in the region of the magnetic poles at spin periods $P \sim 2-3 \mbox{ s}$ by no means implies that the electrodynamic processes become unimportant. Thus, for example, it is well known that the secondary plasma cannot be produced in the region of closed magnetic field lines only at periods $P>P_{cr} \sim 10^5 B_{12}^{4/3} \mbox{ s}$ \citep{IM1995}, where $B_{12}=B/10^{12} \mbox { G}$. Therefore, at $P<P_{cr}$ some of the closed magnetic field lines still remain filled with secondary electron-positron plasma. As we see, the critical period is long enough (more than a month), so the electrodynamic processes in extinct radio pulsars play a significant role for quite a long time. Accordingly, particles can be effectively accelerated in the region of open magnetic field lines where a strong longitudinal electric field is generated. This acceleration results in observable radiation associated with curvature losses.

Clearly, the total intensity of this radiation is related to the number of charged particles that fall into the region of strong longitudinal electric fields. This, in turn, directly depends on the work function of particle escape from the neutron star surface. Thus, analyzing the radiation from extinct radio pulsars, we can draw a definite conclusion about the particle-escape work function and, hence, obtain independent information on the structure of the particle acceleration region near the magnetic poles of radio pulsars.

In this paper we discuss two basic models: the model with hindered particle escape from the neutron star surface \citep{RS1975} and the model with free particle escape (\citealt{Arons1981}; \citealt{Mestel1999}). We show that in the scope of model with hindered particle escape, the number of particles leaving the pulsar magnetosphere is small. So, their radiation cannot be detected with currently available instruments. At the same time, for the model with free particle escape, the number of particles and, hence, the total energy release are large enough for such extinct radio pulsars to be detectable with GLAST or AGILE satellites. Also, within the scope of free particle escape model, extinct radio pulsars can possibly be among the unidentified sources from the 3rd EGRET Catalog.

\section{Model with hindered particle escape from the pulsar surface}

The simplest model we use to estimate the radiation from extinct radio pulsars is the model with hindered particle escape from the pulsar surface. The effects of general relativity for this model do not lead to any significant corrections, because they do not qualitatively change the structure of the electrodynamic equations in the particle-acceleration region \citep{Beskin1999}. Since the particle-escape work function for a surface of a neutron star in the Ruderman-Sutherland model is large, particles only fill the equatorial regions with a strongly curved magnetic field, in which secondary particles can still be produced. As a result, the longitudinal electric field outside the plasma (in particular, on open magnetic field lines) can be strong enough for effective particle acceleration. In this case, the polar cap size is determined by the last closed magnetic field line on which secondary plasma can still be produced \citep{IM1995}:
\begin{equation}
R_{\bot} \approx R\left( \frac{P}{P_{cr}} \right)^{3/8},
\label{1}
\end{equation}

\noindent Here $R$ is a radius of a neutron star.

Thus, the observed radiation of extinct radio pulsars is determined solely by the primary particles radiation. At that, the same curvature radiation mechanism as for the normal radio pulsars can be considered. Recall, that curvature radiation is similar to synchrotron radiation, but the curvature radius a magnetic field line $R_c$ is used instead of the Larmor radius $r_L=m_ec^2\gamma/eB$ \citep{Zheleznyakov1977}. However, in contrast to young neutron stars, the condition for the production of secondary electron-positron pairs in the polar regions of extinct radio pulsars is violated. Therefore, in case of extinct radio pulsars the radiation should be expected only in high-energy part of the electromagnetic spectrum. We stress that this point distinguishes our analysis from other calculations of gamma-ray emission of normal radio pulsars, for which the entire secondary-particle spectrum should be taken into account (see \citet{Harding2002} and references therein).

We estimate expected flux from radio-quiet pulsars as the product of the energy losses by a single particle and the number of particles leaving the pulsar magnetosphere:
\begin{equation}
F=\frac{\Delta {\cal E} \dot N}{4 \pi d^2}.
\label{2}
\end{equation}

\noindent Here, $\dot N = dn/dt$ is the number of particles that leave a pulsar magnetosphere, $\Delta {\cal E}$ is the energy losses by each of these particles, and $d$ is the distance between a pulsar and an observer. Since the radiation mechanism in this case is curvature radiation, the particle energy losses can be defined as $\Delta {\cal E}$. Here \citep{Zheleznyakov1977})
\begin{equation}
\frac{d {\cal E}}{dt}=\frac{2e^2c}{3R_c{}^2}\gamma^4
\label{3}
\end{equation}

\noindent is the power of the curvature radiation, $R_c \approx R^2/R_{\bot}$ is the curvature radius of magnetic field lines, and  $R_{\bot}\approx 10^4 \mbox{ cm}$ (Eq.~\ref{1}) is the distance between the last closed magnetic field line and the magnetic dipole axis.

It is well known that a strong longitudinal electric field exists only near the neutron-star surface. At rather small distance from a neutron star surface the longitudinal electric field drops quickly even if there is no plasma flowing away in this region \citep{MT1990}. Thus, we should consider two possible scenarios: the particle emits all its kinetic energy near the neutron star surface or it leaves the region of strong longitudinal electric fields virtually not losing its energy. In the former case, the particle energy losses $\Delta {\cal E}$ are determined by the ''radiation time'' $\tau_{\mbox{rad}}$, and $\Delta {\cal E}=mc^2 \gamma$. In the latter case, the energy losses are defined by the ''escape time'' (the time that a particle needs for leaving the region of the strong longitudinal electric fields). Generally, the minimum time $\Delta t=min(\tau_{\mbox{esc}}, \tau_{\mbox{rad}})$ can be taken as $\Delta t$. The escape time is $\tau_{\mbox{esc}}={R_c}/{c} \approx 10^{-3} \mbox{ s}$. For the radiation time we can write the following expression $\tau_{\mbox{rad}}=mc^2 \gamma/(d {\cal E}/dt)$, where the Lorentz factor of the accelerated particles can be estimated as
\begin{equation}
\gamma \sim \frac{eER_\bot}{m_ec^2} \sim \frac{e(4\pi
\rho_{GJ}R_{\bot})R_\bot}{m_ec^2} \sim \frac{4\pi eR_\bot{}^2 B_0}{Pm_ec^3},
\label{4}
\end{equation}

\noindent Here, $B_0$ is a magnetic field on a neutron star surface. As a result, the radiation time can be written as
\begin{equation}
\tau_{\mbox{rad}} \sim \frac{R^4c^5}{\Omega^3\omega_B^3r_eR_{\bot}^8},
\label{5}
\end{equation}

\noindent where $\rho_{GJ}=en_{GJ}=-{\bf B \Omega}/(2\pi c) = B/(Pc)$ is the Golreich-Julian density, $r_e = e^2/m_ec^2$ is the classical electron radius, and $\omega_B = eB/m_ec$ is the cyclotron frequency. Here we want to note that the radiation and escape times are of the same order: $\tau_{\mbox{rad}} \approx 10^{-4} B_{12} \mbox{ s}$, $\tau_{\mbox{esc}}=
R_c/c \approx 10^{-3} \mbox{ s}$. Thus, within the accuracy of our estimate we can use either of these times. Finally, given the particle Lorentz factor, the characteristic energy of the curvature photons can be estimated as
\begin{equation}
E_{ph} \approx 2.9 \times 10^3 {B_{12}}^{-1/2} P^{-3/8} \mbox{ MeV.}
\label{6}
\end{equation}

\noindent We see that the expected energies are within the hard gamma-ray spectrum.

As we mentioned above, within the scope of hindered particle escape model the particle production in the open field lines region is not possible. Still primary particles can appear in the region of strong longitudinal electric field. Here are two possible ways for this region to have a number of primary particles. Firstly, the primary particle production can be associated with diffuse Galactic gamma-ray radiation, which causes the single-photon conversion of photons into electron-positron pairs. In this case the number of particles being produced near the magnetic poles can be estimated as $2 \times 10^5 {\mbox{ particles}}/{\mbox{s}}$ \citep{ShukreRadha1982}. Hence, the expected flux (Eq.~\ref{2}) is
\begin{equation}
F \approx 0,3 \times 10^{-25} d_{10}^{-2}\, {\mbox{ MeV}/ \mbox{cm}^{2} \mbox{s}},
\label{7}
\end{equation}

\noindent where $d_{10} = d/10 {\mbox{ pc}}$.

Another source of primary particles is a diffusion from the region of closed magnetic field lines filled with secondary plasma. The number of particles leaving the pulsar magnetosphere can be estimated as \citep{IM1995}
\begin{equation}
\dot N=\frac{dn}{dt}=-n_{GJ} \gamma R^3 \frac{\Omega^2}{\omega_B}.
\label{8}
\end{equation}

As a result, the number of particles that diffuse into the region of a strong longitudinal electric field can be estimated as
\begin{equation}
\dot N=\frac{4\pi^2 m_e \gamma R^3}{P^3 e^2} \approx 4 \cdot 10^{18} P^{-4}\: \frac{\mbox{particles}}{\mbox{s}},
\label{9}
\end{equation}

\noindent The expected flux from the object (Eq.~\ref{2}) is
\begin{equation}
F \approx 10^{-13} d_{10}^{-2} P^{-4} {\mbox{ MeV}}/{{\mbox{cm}^2 \mbox{s.}}}
\label{10}
\end{equation}

Thus, providing a validity of the model with hindered particle escape, the radiation intensity of extinct radio pulsars turns out to be so low that the radiation power is not enough for the signal of neutron star to be detectable even at $P\sim 2-4 \mbox{ s}$. The main reason is that the number of primary particles that fall into the region of a strong longitudinal electric field is very small, i.e., because the particle escape from the neutron star surface is hindered.

\section{Model with free particle escape from the pulsar surface}

In this section we consider the model with free particle escape from the neutron star surface. Within this model there is only a slight difference between the charge density of the particles flowing away from the polar cap and the Goldreich-Julian density. General relativity effects become significant near the neutron star surface (\citealt{MT1990}; \citealt{Beskin1990}). Since no particles are produced, the model of \citet{Mestel1999}, in which no secondary plasma is generated, is appropriate in this case. The other model \citep{Arons1981} cannot be realized for extinct radio pulsars because it requires a reverse flow of secondary particles (for more details, see \citealt{Beskin1999}). Here a short explanation should be done. The Mestel's model presumes that none of the magnetic field lines near the pulsar surface are preferred because of the effects of general relativity. Therefore the particles cannot be accelerated in this region \citep{Beskin1990}. In other words, the longitudinal electric field that arises from the mismatch between the plasma charge density and the Goldreich charge density decelerates rather than accelerates plasma particles.

Nevertheless, the absence of regular acceleration does not imply that the plasma cannot fill the polar regions. Only effective particle acceleration is impossible. On the other hand, even at small distances from the pulsar surface $r > r_0$, where
\begin{equation}
r_0 \approx 1,8 \cdot 10^6 \left(\frac{M}{M_{\odot}}\right) P^{1/7}
\mbox{ cm},
\label{11}
\end{equation}

\noindent the general relativity effects become negligible. At such distances the difference between the plasma charge density and the Goldreich charge density results in particle acceleration, at least on the half of the polar cap where the magnetic field lines deviate from the spin axis (i.e., those magnetic field lines for which the angle between the magnetic field line and the spin axis increases with distance from the neutron star). Below, we assume that particle acceleration begins only after the radius $r = r_0$ is reached. At smaller radii plasma on open magnetic field lines rotates rigidly with a neutron star. Finally, recall that in Mestel's model, as in Arons's model, the number of particles leaving the magnetosphere of the neutron star is close to the Goldreich-Julian density near the polar-cap surface, i.e.
\begin{equation}
\dot N = \pi R_0^2 n_{GJ} c \approx \frac{10^{30}
{\Omega}^2}{c|e|} \mbox{ particles/s}.
\label{11'}
\end{equation}

\noindent Here, $R_0=(\Omega R/c)R$ is the polar cap radius.

To define the electric potential, we write the Poisson equation in a rotating coordinate system \citep{Mestel1999}
\begin{equation}
\label{12} \Delta \varphi=4\pi (\rho_{GJ}-\rho_e).
\end{equation}

\noindent Here, the Goldreich-Julian charge density and the particle charge density can be written as
\begin{eqnarray}
\rho_{GJ}  = -\frac{\vec \Omega \vec B}{2\pi c}=
\frac{\Omega B}{2\pi c}\cos \theta_b ,\\
\rho_e  = -\frac{\Omega B}{2\pi c}\cos \theta_b^{(0)},
\end{eqnarray}

\noindent where $\theta_b$ is the current angle between the magnetic field line and the spin axis, and $\theta_b^{(0)}$ is the angle between the magnetic field line and the spin axis at the bottom of the acceleration region $r = r_0$.

To simplify our calculations, we consider axisymmetric case of an aligned rotator, for which all magnetic field lines are preferred. Using the expressions for the plasma charge density and the Goldreich charge density, we can write the following equation
\begin{equation}
\Delta \varphi=\frac{9}{4} \frac{\Omega B(r_0)}{c} \left
(1-\frac{r_0}{r}\right) {\left (\frac{r_0}{r}\right)}^3 \sin^2 \theta
\label{15}
\end{equation}

\noindent The boundary conditions can be defined as
\begin{equation}
\varphi \left( r=r_0 \right)=0; \hspace{2em} \varphi' \left( r=r_0 \right)=0; \hspace{2em} \varphi \left(r=R_L \sin^2\theta \right)=0.
\label{16}
\end{equation}

\noindent Here, $R_{\rm L} = c/\Omega$ is the radius of the light cylinder. As a result, for a zero approximation we have
\begin{equation}
\varphi \approx \frac{\Omega B_0}{c} \frac{R^3}{R_L},
\label{17}
\end{equation}

\noindent Thus,
\begin{equation}
\gamma \approx \frac{e \varphi}{m_ec^2}=\left(\frac{\Omega R}{c}\right)^2 \left(\frac{\omega_B R}{c}\right).
\label{18}
\end{equation}

In the above expressions we used the fact that magnetic flux should be constant (''frozen'' magnetic field lines), which yields $B(r_0)r_0{}^3=B_0 R^3$. The radiation and escape times can be estimated as $\tau_{\mbox{rad}}$ and $\tau_{\mbox{esc}}$, the same way as for the model with hindered particle escape from the neutron star surface. Within the model with free particle escape $\tau_{\mbox{rad}} \approx \tau_{\mbox{esc}}$ at $P =P_{\mbox{break}}$, where
\begin{equation}
P_{\mbox{break}} = \frac{2\pi R}{c} \left(\frac{2r_e}{3R}\right)^{1/7}
\left(\frac{\omega_B R}{c}\right)^{3/7} \approx 0.95 \mbox{ s}.
\label{19}
\end{equation}

\noindent  As we see, the break takes place at spin periods of greatest interest -- when a neutron star just crossed the death line.

The estimated energy of the curvature gamma-ray photons leaving the magnetosphere of the neutron star $E_{\rm ph}=\hbar c \gamma^3/R_c
\approx 8 \cdot 10^{-5} P^{-13/2} \mbox{\rm{ erg}} \approx 0,05P^{-
13/2} \mbox{\rm{ GeV}}$ is within the gamma-ray range. Finally, the radiation intensity can be estimated as
\begin{eqnarray}
W \approx m_ec^2 \gamma \dot N \approx \left (\frac{\Omega R}{c}\right)^4
\left(\frac{\omega_B R}{c}\right)^2  \frac{m_ec^3}{r_e} \approx 5.8 \cdot
10^{31} \frac{B_{12}^2}{P^4}\: \frac{\mbox{erg}}{\mbox{s}}, \hspace{1em} \\
\mbox{if }P < P_{\mbox{break}} \qquad \qquad \nonumber \hspace{16em} \\
W \approx \frac{2e^2 c}{R_c{}^2} \gamma^4 \dot N \approx
\left(\frac{\Omega R}{c}\right)^{11} \left(\frac{\omega_B R}{c}
\right)^5 \frac{2m_ec^3}{3R} \approx 4 \cdot 10^{31}
\frac{B_{12}^5}{P^{11}} \: \frac{\mbox{erg}}{\mbox{s}}, \\
\mbox{if } P > P_{\mbox{break}}.\nonumber \hspace{20em}
\end{eqnarray}

\noindent The flux from such neutron stars near the Earth can be defined by the following expression
\begin{equation}
F \approx \frac{W_1}{\pi \alpha_0^2 d^2} \frac{\mbox{erg}}{\mbox{cm}^2 \mbox{s}},
\end{equation}

where $d$ is the distance to a neutron star, and $\alpha_0=R_0/R$ is defined by the polar cap size $R_0$. Thus, for an oblique rotator the averaged flux from an extinct radio pulsar is
\begin{equation}
F_{normalized}=\frac{2R_0}{2 \pi R \sin \chi} \frac{F}{E_{ph}} \frac{\mbox{photons}}{\mbox{cm}^2 \mbox{s}}
\end{equation}

Here $\chi$ is the angle between the magnetic and spin axes. As a result, the expected radiation fluxes from extinct radio pulsars within the scope of free particle escape model are

\begin{eqnarray}
\nonumber F_1 \approx 6 \cdot 10^{-4} B_{12}^{5/13}R_6^{12/13} d_{pc}^{-2} E_{ph}^{-6/13} \quad \frac{\mbox{photons}}{{\mbox{cm}^2 \mbox{s}}}, \\
P < P_{\mbox{break}},\nonumber
\\
\nonumber F_2 \approx  8 \cdot 10^{-2} B_{12}^{2/13}R_6^{10/13} d_{pc}^{-2} E_{ph}^{8/13} \quad \frac{\mbox{photons}}{\mbox{cm}^2 \mbox{s}}, \\
P > P_{\mbox{break}}.
\end{eqnarray}

Since the expected fluxes are large, we conclude that for the distances $d \sim 100-200 \mbox{ pc}$ characteristic for the nearest neutron stars, such fluxes can most likely be detected with GLAST and AGILE satellites.

On the other hand, since the radiation intensity strongly depends on the spin period of the neutron star, radio-quiet pulsars with long spin periods (i.e., those that have long ceased to radiate in the radio band or those with comparatively long spin periods at the stage of radio emission) can no longer be detected. So there are three conditions under which extinct radio pulsars are detectable: the validity of the model with free particle escape from the neutron star surface, relatively small distances to such stars, and short spin periods of these stars.

Figure~\ref{sensitivity06extinct_B_12} compares the threshold sensitivities of the modern gamma-ray missions (particularly, GLAST, AGILE, and EGRET) and the photon count rates for the model with free particle escape from the pulsar surface. We assumed the distance from a neutron star to the observer is 100 pc, radius a star is 10 km, and magnetic field is $10^{12}$ G. As we see, the closest extinct pulsars (if any of them exist in such a vicinity) can be detectable by the future missions such as GLAST and AGILE.

\section{Conclusions}

Thus, we have shown that extinct radio pulsars can be observed for some time as intense gamma-ray sources. For nearby neutron stars, they can be detected with the receivers installed on the modern gamma-missions.

On the other hand, since the dependence on the spin period $P$ of
neutron star in the expressions for the radiation intensity is very strong,
radio-quiet pulsars with sufficiently long spin periods (i.e., those
which have long ceased to exist as radio pulsars or those for which
the spin period was relatively long even at the radio emission
stage) still cannot be detected. Therefore, we can formulate three
conditions for the possibility of detecting gamma-ray emission from
extinct radio pulsars: a validity of the model with free particle
escape, a relatively small distance to such stars, and short spin periods. Providing all three conditions are satisfied, the
detection of gamma-ray emission with the parameters described above
would be a direct evidence for the free particle escape from the
neutron star surface.

\section*{Acknowledgments}
We thank S.B. Popov andM.E. Prokhorov
for fruitful discussions. This study was supported by
the Russian Foundation for Basic Research (project
no. 02-02-16762).

{\onecolumn
\begin{figure}
\centering
\includegraphics[width=15cm,height=12cm,keepaspectratio]{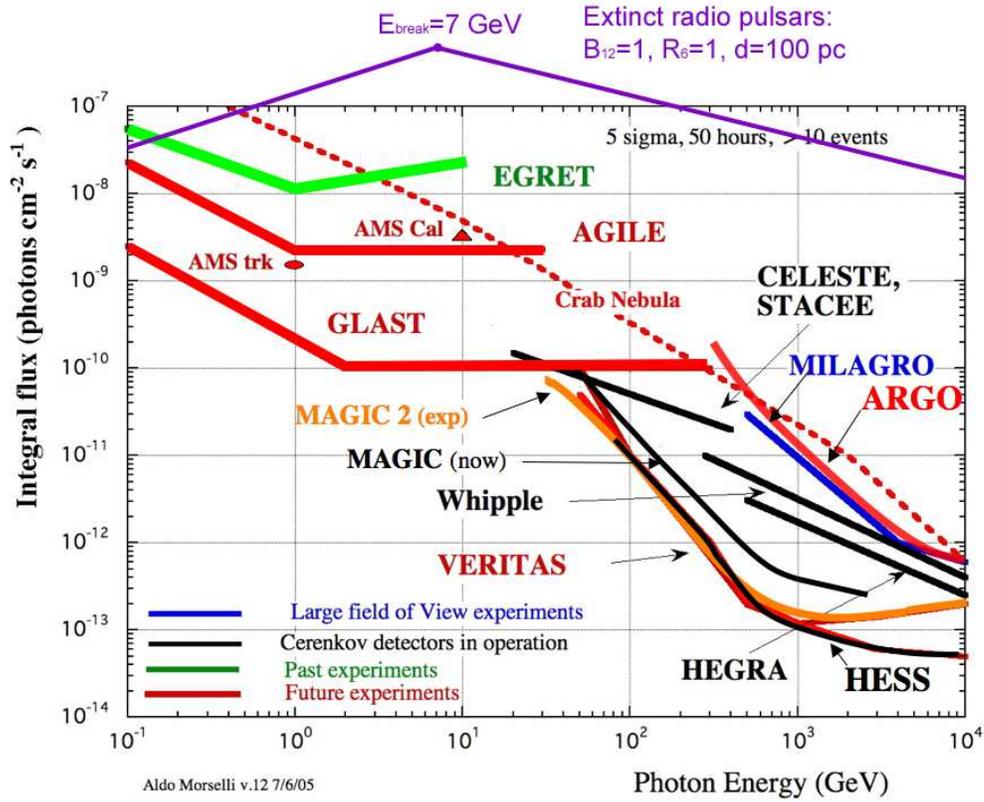}
\caption{\small{Comparison of the sensitivities of present and future detectors in the
gamma-ray astrophysics \citep{Morselli} and the predicted radiation intensities of extinct radio pulsars.}}
\label{sensitivity06extinct_B_12}
\end{figure}
}

\end{document}